\documentclass{ws-procs9x6}

\begin{document}

\title{Uncertainty relation \\
for quantum gravity
\footnote{\uppercase{B}ased on the talk at the 4th 
\uppercase{V}.\uppercase{N}.\uppercase{G}ribov \uppercase{M}emorial 
\uppercase{W}orkshop (gribov85), \uppercase{J}une, 17-20, 2015, 
\uppercase{C}hernogolovka, \uppercase{R}ussia. \uppercase{T}o be published 
in the \uppercase{P}roceedings. }
}

\author{Y\lowercase{a}.~I. AZIMOV }

\address{Petersburg Nuclear Physics Institute, \\National Research Center
``Kurchatov Institute'', \\
Gatchina, 188300, Russia \\
E-mail: azimov@thd.pnpi.spb.ru}

\maketitle

\abstracts{
Discussion of physical realization of coordinates demonstrates
that the quantum theory of gravity (still absent) should be
non-local and, probably, non-commutative as well.}

Various fields became familiar objects for contemporary physics.
But their description may be different. Classical fields are
usually described as numerical functions $A(x,y,z,t)$ of
coordinates and time (and, m.b., of some other parameters).
It is implicitly assumed that both the coordinates, time, and
the field values can be fixed with arbitrary good precision.
The above approach is applicable, in particular, to classical
gravitation field.

Quantum fields are generally described as operator functions
$A(x,y,z,t)$, also depending on coordinates and time. Usually,
the field operator $A$ may be diagonalized. And it is assumed
again that both the coordinates and time, and the diagonal
field values can be fixed with arbitrary good precision.

Quantum theory of gravitation is still absent, though the literature
contains many models and attempts to construct it (see, {\it e.g.},
the recent reviews~\cite{jl,aa,bs}). Many of their authors act usually
in the familiar way and try to formulate the quantum gravity as
a local field theory. Here we discuss, whether this is possible.

Having a particular coordinate frame, one may physically mark a
particular physical point by means of a "point-like" test-body being
at rest in the frame. By definition, the test-body should be chosen so
to minimize its influence on the external fields. This can certainly
be realized in the classical case or for the non-relativistic
quantum case. In both cases, space coordinates can be fixed with
arbitrary precision.

The situation looks different for the relativistic quantum case
accounting for the finite velocity of light $c$. Position of a
test-body of mass $M$, even if it is point-like, can not be fixed
with better precision than its Compton length~\cite{ss}
\begin{equation}
\lambda_{\,C} = \frac{\hbar}{M\,c}\,\,.
\label{comp}
\end{equation}
For most fields, this fact does not restrict precision of coordinates:
for a better precision one should use a heavier test-body. However,
if the gravitation field is switched on, one encounters the
inconsistency: a heavier test-body provides a better precision for
its position, but stronger distorts the external gravitation field.

The same conclusion may be obtained in other way(s). If consideration
begins with gravitation field in non-relativistic case and/or without
quantum effects, then one could take a test-body to have very small mass,
to not perturb the external field. However, accounting for both relativistic
and quantum effects in this situation generates large
uncertainty in the position of such test-body.

Let us consider the situation in some more detail. To fix the space
point $(x, y, z)$ up to the sphere of the radius $\delta_r$ one should
use a test-body with the mass
\begin{equation}
M_\delta\approx \frac{\hbar}{\delta_r\, c}\,\,.
\label{mass}
\end{equation}
Then, to the gravitation field $\varphi$ near the point $(x, y, z)$,
in the Newton approximation, the test-body appends the uncertainty
\begin{equation}
\delta_\varphi\approx \frac{G\,\hbar}{\delta_r^2 \,c}\,\,,
\label{field}
\end{equation}
where $G$ is the Newton constant. This uncertainty relation between
coordinates and gravitation field may be rewritten as
\begin{equation}
\left(\frac{\delta_\varphi}{c^2}\right)\,\left(\frac{\delta_r}
{\lambda_P}\right)^2
\approx 1\,\,,
\label{varf}
\end{equation}
with $\lambda_P=\sqrt{G\hbar/c^3}$ being the Planck length.
Note that the uncertainty relation (\ref{field}) disappears if
at least one of the three conditions is satisfied: non-quantum
case ($\hbar\to 0$); non-relativistic case ($c\to\infty$);
absence of gravitation ($G\to0$).

Let us discuss meaning of these relations. When gravity is
present, the test-body of mass $M$ generates, in addition to
the Compton length (\ref{comp}), one more length scale, the
Schwarzschild radius~\cite{LL}
\begin{equation}
\lambda_{\,S}=\frac{2\,GM}{c^2} \,\,.
\label{sch}
\end{equation}
At small $M$, $\lambda_{\,S}<\lambda_{\,C}$. But at $M>M_P/\sqrt{2}$,
where
\begin{equation}
M_P=\sqrt{\frac{\hbar\,c}{G}}
\label{pl}
\end{equation}
is the Planck mass, the inequality changes its sign. If $\lambda_{\,S}$
exceeds the proper dimension of the body, then it corresponds to the
radius of the Schwarzschild horizon, which internal area is invisible
for the frame, where the body is at rest. Therefore, it is $\lambda_{\,S}$
that determines the position uncertainty for a heavier test-body. The
minimal physical uncertainty is achieved when
\begin{equation}
\lambda_{\,S}=\lambda_{\,C}=\sqrt2\,\lambda_{\,P}\,\,.
\label{min}
\end{equation}
Thus, if quantum theory describes gravitation by a field, the field
should be non-local, since its coordinates can never be physically fixed
with precision better than $\sqrt2\,\lambda_{\,P}$.

Furthermore, the combination $\varphi/c^2$ is just that appearing
in the metric tensor~\cite{LL}. Therefore, Eq.(\ref{varf}) means that
uncertainty of coordinates is related to uncertainties in the metric
tensor and, may be, in other coordinates (and/or time). This may
hint at non-commutativity of operators for different space-time
coordinates in the future quantum gravity theory.

Such properties are not necessary for quantum theory  of other known
interactions in absence of gravity (though might work as well in some
particular cases). This is related to a unique feature of gravitation:
in contrast to any other interactions, there can not exist a test-body
which would be neutral (sterile) in respect to gravity.

\section*{Acknowledgments}
I thank I.T. Dyatlov for useful discussion. This work is supported by
the Russian Science Foundation (Grant No.14-22-00281).

\end{document}